\input ./style/arxiv-ba.cfg
\documentclass[ba,linksfromyear,preprint]{imsart}
\makeatletter
   \@ifpackageloaded{natbib}{}{\usepackage{natbib}}
\makeatother

\pubyear{2015}
\volume{10}
\issue{4}
\firstpage{965}
\lastpage{990}
\doi{10.1214/14-BA927}

\startlocaldefs
\newcommand{\enquote}[1]{``#1''}

\newtheorem{theorem}{Theorem}

\newcommand{\bfTheta} {\boldsymbol{\Theta}}
\newcommand{\bfSigma} {\boldsymbol{\Sigma}}

\newcommand{\bfI} {\mathbf{I}}
\newcommand{\bfY} {\mathbf{Y}}
\newcommand{\bfX} {\mathbf{X}}
\newcommand{\bfE} {\mathbf{E}}
\newcommand{\bfW} {\mathbf{W}}
\newcommand{\bfD} {\mathbf{D}}
\newcommand{\bfB} {\mathbf{B}}
\newcommand{\bfK} {\mathbf{K}}
\newcommand{\bfU} {\mathbf{U}}
\newcommand{\bfPhi} {\mathbf{\Phi}}
\newcommand{\bfx} {\mathbf{x}}
\newcommand{\bfy} {\mathbf{y}}
\newcommand{\bfu} {\mathbf{u}}

\renewcommand{\Pr}{\mathsf{Pr}}
\newcommand{\E}{\mathsf{E}}

\newcommand{\normal}{\mathsf{N}}
\newcommand{\MN}{\mathsf{MN}}
\newcommand{\Poi}{\mathsf{Poi}}
\newcommand{\GWis}{\mathsf{Wis}}
\newcommand{\cone}{\mathsf{P}}

\newcommand\independent{\protect\mathpalette{\protect\independenT}{\perp}}
\def\independenT#1#2{\mathrel{\rlap{$#1#2$}\mkern2mu{#1#2}}}

\endlocaldefs

\begin{document}

\begin{frontmatter}
\title{Restricted Covariance Priors with Applications in Spatial Statistics}
\runtitle{Restricted Covariance Priors with Applications in Spatial Statistics}

\begin{aug}
\author[addr1]{\fnms{Theresa R.} \snm{Smith}\corref{}\ead[label=e1]{tsmith7@uw.edu}},
\author[addr2]{\fnms{Jon} \snm{Wakefield}\ead[label=e2]{jonno@uw.edu}},
\and
\author[addr3]{\fnms{Adrian} \snm{Dobra}\ead[label=e3]{adobra@uw.edu}}

\runauthor{T. R. Smith, J. Wakefield, and A. Dobra}

\address[addr1]{Department of Statistics, University of Washington,
Seattle, WA 98195, \printead{e1}}

\address[addr2]{Departments of Statistics and Biostatistics,
University of Washington, Seattle, WA 98195,\\ \printead{e2}}

\address[addr3]{Departments of Statistics,
Biobehavioral Nursing, and Health Systems and the Center for
Statistics and the Social Sciences, University of Washington,
Seattle, WA 98195, \printead{e3}}
\end{aug}

%
\begin{abstract}
We present a Bayesian model for area-level count data that uses
Gaussian random effects with a novel type of G-Wishart prior on the
inverse variance--covariance matrix. Specifically, we introduce a new
distribution called the truncated G-Wishart distribution that has
support over precision matrices that lead to positive associations
between the random effects of neighboring regions while preserving
conditional independence of non-neighboring regions. We describe Markov
chain Monte Carlo sampling algorithms for the truncated G-Wishart prior
in a disease mapping context and compare our results to Bayesian
hierarchical models based on intrinsic autoregression priors. A
simulation study illustrates that using the truncated G-Wishart prior
improves over the intrinsic autoregressive priors when there are
discontinuities in the disease risk surface. The new model is applied
to an analysis of cancer incidence data in Washington State.
\end{abstract}

%

\end{frontmatter}


\section[Introduction]{Introduction}
Spatial data arise when outcomes and predictors of interest are
observed at particular points or regions inside a defined study area.
Spatial data sets are common in many fields including environmental
science, economics, and epidemiology. In epidemiology, understanding
the underlying spatial patterns of a disease is an important starting
point for further investigations. The risk of disease inherently varies
in space because the risk factors are non-uniformly distributed in
space. Such risk factors may include lifestyle variables such as
alcohol and tobacco use or exposure levels of environmental causes of
disease such as air pollution or UV radiation. We expect that these
risk factors are positively correlated in space meaning that nearby
areas will have similar exposure levels or underlying characteristics.
That is, we assume risk factors obey Tobler's first law of geography:
``everything is related to everything else, but near things are more
related than distant things'' \citep{tobler1970computer}.

In many studies, underlying disease risk factors are unknown or
unmeasured. Bayes\-ian models account for unknown or unmeasured risk
factors using priors chosen to mimic their correlation structure. The
most common Bayesian framework for area-level spatial data uses
Gaussian random effects with a covariance structure that imposes
positive spatial dependence between random effects of neighboring or
near-by areas \citep{besag1991bayesian,diggle1998model,banerjee2004hierarchical}. The non-Gaussian spatial
clustering and Potts model based priors also impose positive dependence
in the relative risks of neighboring areas \citep{knorr2001shared,green2002hidden}. More recently, several authors have developed
modifications to existing models, specifically to preserve positive
dependence for spatial statistics applications \citep{wang2013class,hughes2013dimension}. Further, positive spatial dependence is usually
imposed in geostatistical models for data observed point-wise rather
than area-wise. For example, the Mat\'{e}rn family of marginal
covariance functions for Gaussian random fields yields positive
correlations between observations at two locations locations, with the
magnitude of the correlation decreasing with distance \citep
{stein1999interpolation,diggle2007model}.

We present a Bayesian model for area-level count data that uses
Gaussian random effects with a novel type of G-Wishart prior on the
inverse variance--covariance matrix. The usual G-Wishart or hyper
inverse Wishart prior restricts off-diagonal elements of the precision
matrix to 0 according to the edges in an undirected graph \citep
{dawid1993hyper,roverato2002hyper}. \citet{dobra2011bayesian} use the
G-Wishart prior to analyze mortality counts for ten cancers in the
United States using a Bayesian hierarchical model incorporating
Gaussian random effects with a separable covariance structure. Their
comparisons show that allowing different strengths of association
between paris of neighboring states can have advantages over
traditional conditional autoregressive priors that assume the same
strength of conditional association across the study region. However,
the G-Wishart prior allows for both positive and negative conditional
associations between neighboring areas.

The truncated G-Wishart distribution that we introduce only has
support over precision matrices that lead to positive conditional
associations. We describe Markov chain Monte Carlo (MCMC) algorithms
for this new prior and construct a Bayesian hierarchical model for
areal count data that uses the truncated G-Wishart prior for the
precision matrix of Gaussian random effects. We show via simulation
studies that risk estimates based on a model using the truncated
G-Wishart prior are better than those based on conditional
autoregression when the outcome is rare and the risk surface is not smooth.
For univariate data, there is little information to identify the
parameters of the spatial precision matrix; however, we can share
information across outcomes in
a multivariate model by assuming a separable covariance structure. We
illustrate the improvement of using the truncated G-Wishart prior in a
separable model (measured via cross-validation) using cancer incidence
data from the Washington State Cancer Registry.

The structure of this paper is as follows. In Section 2, we present
our modeling framework and give a brief overview of conditional
autoregressive models. In Section~3, we define the truncated G-Wishart
distribution and give the details of an MCMC sampler for estimating
relative risks in a spatial statistics context. In Section 4, we
present a simulation study based on univariate disease mapping using
the geography of the counties of Washington State. Finally, in Section
5, we extend the univariate truncated G-Wishart model to multivariate
disease mapping using the separable Gaussian graphical model framework
of \cite{dobra2011bayesian}.
\section[Background]{Background}
\subsection{Notation}
Let $\mathcal{A} = \{A_1, \dots A_n\}$ be a set of non-overlapping
geographical areas, and let $\bfy= \{y_1, \dots, y_n \}$ represent the
set of counts of the observed number of health events in these areas.
Possible health events include deaths from a disease, incident cases of
a disease, or hospital admissions with specific symptoms of a disease.
Next, let $\bfE= \{E_1, \dots, E_n\}$ be the set of expected counts
and $\bfX= \{\bfx_1 \dots\bfx_n\}$ be a matrix where $\bfx_i$ is a
vector of suspected risk factors measured in area $i$. The expected
counts account for differences in known demographic risk factors. If
the population in each area is stratified into $J$ groups (e.g., gender
and 5 year age-band combinations), then the expected count for each
area is
\begin{align}
E_{i} & = {\sum_{j =1}^J q_{j} P_{ij} }, \nonumber
\end{align}
where $P_{ij}$ is the population in area $i$ in demographic group $j$
and $q_{j}$ is the rate of disease in group $j$. The rates $q_{j}$ may
be estimated from the data if the disease counts are available by
strata (internal standardization) or they may be previously published
estimates for the rates of disease (external standardization).

A generic Bayesian hierarchical model for data of this type is:
\begin{align}
y_{i}\mid\bfy_{-i}, {E}_i, \theta_i & \sim\Poi(E_i \theta_i),
\nonumber\\
\log(\theta_i)& = \bfx_i^T \beta+ u_i, \nonumber\\
\pi(\bfu) & = H, \nonumber
\end{align}
where $\bfy_{-i}$ is the vector of counts with area $i$ excluded and
$H$ is a probability distribution with spatial structure. Most choices
of $H$ encode the belief that the residual spatial random effects, $\bfu
$, of nearby areas have similar values. This restriction follows from
the interpretation of the random effects as surrogates for unmeasured
risk factors, which are generally assumed to be positively correlated
in space. The inclusion of $H$ produces smoother (though biased)
estimates of the vector of relative risks, $\boldsymbol{\theta}$, with
reduced variability compared to the maximum likelihood estimates
$\widehat{\boldsymbol{\theta}}=\mathbf{y} /\mathbf{E}$. These maximum
likelihood estimates, called standardized incidence ratios (SIRs) or
standardized mortality/morbidity ratios (SMRs), have large sampling
variances when the expected counts are small. A key task in modeling
areal count data is to choose a prior $H$ that is flexible enough to
adapt to the smoothness of the risk surface.

\subsection{Existing Models for Areal Count Data}
The most common choice for $H$ is the Gaussian conditional
autoregression or CAR prior \citep{besag1974spatial,rue2005gaussian},
which is a type of Gaussian Markov random field. The CAR model for a
vector of Gaussian random variables is defined by a set of conditional
distributions.
The conditional distribution for the random variable, $u_i$, given the
other variables, $\mathbf{u}_{-i}$, is
\begin{align}
u_i\mid\bfu_{-i} & \sim{\normal}\left( \sum_{j:j\neq i} b_{ij}u_j,
\tau_i^{2} \right). \nonumber
\end{align}
The joint distribution of the vector $\bfu$ is a mean-zero multivariate
normal distribution with precision $\bfD^{-1}(\bfI-\bfB)$, where
$B_{ij}=b_{ij}$, $B_{ii} = 0$, and $D_{ii}=\tau^2_i$. This is a proper
joint distribution if $\mathbf{D}^{-1}(\bfI- \mathbf{B})$ is a
symmetric, positive definite matrix \citep{banerjee2004hierarchical}.

The \textit{intrinsic conditional autoregression} or ICAR prior is the
most commonly used prior for spatial random effects within the class of
CAR priors. Under the ICAR prior, the conditional mean for a given
random effect is the weighted average of the neighboring random
effects, and the conditional variance is inversely proportion to the
sum of these weights:
\begin{align}
u_i \mid\bfu_{-i}\sim{\normal}\left( \frac{1}{\omega_{i+}}\sum
_{j:j\neq i} \omega_{ij} u_j, \frac{\tau^2_u}{\omega_{i+}} \right).
\label{ICAR}
\end{align}
Here $\omega_{ij}$ is nonzero if regions $i$ and $j$ are neighbors
(i.e., share a border) and 0 otherwise; $\omega_{i+}$ is the sum of all
of the weights for a specific area. A binary specification for $\bfW= \{
\omega_{ij}; i,j = 1, \dots, n\}$ is frequently used, though other
weights that incorporate the distance between areas can also be used
\citep{white2009stochastic}. In the binary case, $\omega_{ij}=1$ for
neighboring regions and $\omega_{i+} = n_i$, the number of regions that
border area $i$. Under this specification, the conditional mean for a
particular random effect is the average value of the random effects for
the neighboring regions, and the conditional variance is inversely
proportional to the number of neighbors of the area.

\cite{besag1991bayesian} use a CAR prior for spatial random effects in
a disease mapping context in what has become known as the \textit
{convolution model}:
\begin{align*}
\log(\theta_i)= \bfx_i^T\beta+ v_i + u_i.
\end{align*}
Here $v_i$ is a non-spatial random effect and $u_i$ is a spatial random
effect. The prior for $\mathbf{v}$ is $\normal(0,\sigma_v^2 \bfI)$, and
the prior for $\mathbf{u}$ is the ICAR prior.

Though popular, the convolution model has several drawbacks. First,
there are only two parameters ($\sigma^2_v$ and $\tau^2_u$) to control
the level of smoothing with only one of these ($\tau^2_u$) contributing
to the spatial portion of the model. This parsimony is ideal for
estimating a smooth risk surface
in the presence of large sampling variably, which is a common issue for
rare diseases or for small area estimation. However, using ICAR random
effects can lead to over-smoothing, which masks
interesting features of the risk surface, including sharp changes.
Several authors have addressed this issue by incorporating flexibility
in the conditional independence structure of the relative
risks \citep{knorr2000bayesian,green2002hidden,lee2013locally,lee2013bayesian}.
These approaches are fairly parsimonious, but estimating the parameters
requires careful reversible jump MCMC or access to data from previous
years. In contrast, we develop a locally-adaptive approach with a
separate parameter for the strength of spatial association between each
pair of neighboring areas while preserving the conditional independence
structure.

A second drawback is that the ICAR prior is improper. The joint
distribution implied by the conditional specification in (\ref{ICAR})
is a singular multivariate normal distribution with precision matrix
$\tau^2_u(\bfD_{\omega} - \bfW)$, where $\bfD_{\omega}$ is a diagonal
matrix with elements $D_{ii}=\omega_{i+}$. Since each row of $\bfD
_{\omega} - \bfW$ sums to $0$, this precision matrix does not have full
rank, and the joint prior for $\mathbf{u}$ is improper. One way to
alleviate both the over smoothing and the singularity issues is through
the addition of a spatial autocorrelation parameter $\rho$:
\begin{align}
u_i \mid\bfu_{-i}\sim{\normal}\left( \frac{\rho}{\omega_{i+}}\sum
_{j:j\neq i} \omega_{ij} u_j, \frac{\sigma^2_u}{w_{i+}} \right).
\nonumber
\end{align}
This specification is called the proper CAR because it gives rise to a
proper joint distribution as long as $\rho$ is between the reciprocals
of the largest and smallest eigenvalues of $\bfD_{\omega}^{-1/2} \bfW
\bfD_{\omega}^{-1/2}$ \citep{banerjee2004hierarchical}. For the binary
specification of $\bfW$, this always includes $\rho\in[0, 1)$. The
relationship between $\rho$ and the overall level of spatial smoothing
in the proper CAR prior is complex. The prior marginal correlations
between the random effects of neighboring areas increase very slowly as
$\rho$ increases, with substantial correlation obtained only when $\rho
$ is very close to $1$ \citep{besag1995conditional}. Further, as $\rho$
increases, the ordering of these marginal correlations is not fixed
\citep{wall2004close}.

Nonetheless, the ICAR prior remains a popular choice for
spatially correlated errors in many applied settings. The conditional
specification in (\ref{ICAR}) is parsimonious, and
one only needs to specify a single prior for the precision of the
spatial random effects. Prior specification has received some attention
in the literature \citep{fong2009bayesian,sorbye2014scaling}. Further,
off-the-shelf MCMC routines for the ICAR and convolution models are
available in WinBUGS \citep{lunn2000winbugs} and various \texttt{R}
packages. Fast computation of approximate marginal posterior summaries
is available using integrated nested Laplace approximation (INLA) \citep
{rue2009approximate}.

\section[Methods]{Methodology}
An alternative to specifying the prior for spatial random effects based
on a set of conditional distributions is to work directly with the
joint distribution. A Gaussian graphical model or covariance selection
model is a set of joint multivariate normal distributions that obey the
pairwise conditional independence properties encoded by an undirected
graph, $G$ \citep{dempster1972covariance,lauritzen1996graphical}. This
graph has two elements: the vertex set $V$ and the edge list $E$. The
absence of an edge between two vertices corresponds to conditional
independence and implies a specific structure for the precision matrix
of the joint distribution. If $\mathbf{u}$ follows a multivariate
normal distribution with precision matrix~$\bfK$, then $\mathbf{u}$
follows a Gaussian graphical model if $u_i \independent u_j \mid
\mathbf{u}_{V\backslash(i,j)} \iff(i,j) \not\in E \implies K_{ij}= 0
$ for any pairs $i$ and $j$. Here $\mathbf{u}_{V\backslash(i,j)}$ is
the vector $\mathbf{u}$ excluding the $i{\text{th}}$ and $j{\text{th}}$
elements.

The conjugate prior for the precision matrix in the Gaussian setting
is the Wishart distribution, which is a distribution over all
symmetric, positive definite matrices of a fixed dimension. The Wishart
distribution has two parameters. The first is a scaler $\delta> 2$,
which controls the spread of the distribution. The second is an $n
\times n$ matrix $\bfD$, which is related to the location of the
distribution. For $\bfK\sim\GWis(\delta, D)$, $\E(\bfK) = (\delta+n-1)
\bfD^{-1}$ and mode$(\bfK) = (\delta-2)\bfD^{-1}$. The G-Wishart
distribution is the conjugate prior for the precision matrix in a
Gaussian graphical model \citep{dawid1993hyper,roverato2002hyper}. The
G-Wishart distribution is a distribution over $\cone^+(G)$, the set of
all symmetric, positive definite matrices with zeros in the
off-diagonal elements that correspond to missing edges in $G$. The
density of the G-Wishart distribution for a matrix $\bfK$ is
\begin{align}
\Pr(\bfK\mid\delta, \bfD, G) & = \frac{1}{I_1(G, \delta, \bfD)} |\bfK
|^{(\delta-2)/2} \exp\left( -\frac{1}{2} \left<\bfK,\bfD\right>\right)
\mathbf{1}_{\bfK\in\cone^+(G)},\label{gwdistro}
\end{align}
where $\left< A, B\right>$ is the trace of $A^T B$. The normalizing
constant $I_1(G,\delta, \bfD)$ has a closed form when $G$ is a
decomposable graph and can be estimated for general graphs using the
Monte Carlo method proposed by \cite{atay2005monte}.

\subsection{Truncated G-Wishart Distribution}
We propose a new G-Wishart distribution called the truncated G-Wishart
distribution that imposes additional constraints on $\bfK$. This is a
distribution over positive definite matrices where the off-diagonal
elements that correspond to (non-missing) edges in $E$ are less than
$0$. This restriction means that all pairwise conditional (or partial)
correlations are positive because
\begin{align}
\mathsf{cor}\left(u_i, u_j \mid\bfu_{V\backslash\{i,j\} } \right)& =
\frac{-K_{ij}}{\sqrt{K_{ii} K_{jj}}}. \nonumber
\end{align}
This restriction is attractive in a spatial statistics context where we
believe neighboring areal units are likely to be similar to each other,
given the other areas.

If $\bfK$ follows a truncated G-Wishart distribution, then
\begin{align}
\Pr(\bfK\mid G, \delta, \bfD) & = \frac{1}{I_2(G, \delta, \bfD)} |\bfK
|^{(\delta-2)/2} \exp\left( -\frac{1}{2} \left<\bfK, \bfD\right>\right
) \mathbf{1}_{\bfK\in\cone^+(G) \cap\mathcal{S}^0}. \label{ngwishdistro}
\end{align}
Here $I_2(G,\delta, \bfD)$ is the unknown normalizing constant, and
$\mathcal{S}^0$ is the set of matrices with negative off-diagonal
elements. The normalizing constant in (\ref{gwdistro}) is finite as
long as $\delta> 2$ and $\bfD^{-1} \in\cone^+(G)$ \citep{atay2005monte}. The normalizing constant in (\ref{ngwishdistro}) is
finite under the same conditions because the support of the truncated
G-Wishart is a subset of the support of the G-Wishart distribution.
The mode of the truncated G-Wishart is again $(\delta-2)\bfD^{-1}$, and
for this reason we only consider $\bfD^{-1} \in\cone^+(G) \cap\mathcal
{S}^0$. In this paper, we write $\mathsf{TWis}_G$ for the truncated
G-Wishart distribution and $\mathsf{Wis}_G$ for the G-Wishart
distribution.\eject

\cite{atay2005monte} and \cite{dobra2011bayesian} transform $\bfK$ to
the Cholesky square root, which we call $\bfPhi$, because it is easier
to handle the positive definite constraint in the transformed space.
In the G-Wishart case, the elements of $\bfPhi$ are either variation
independent or are deterministic functions of other elements. We call
the off-diagonal elements of $\bfPhi$ that correspond to missing edges
in the graph $G$ ``non-free.'' These are deterministic functions of the
``free'' elements: the diagonal elements and the off-diagonal elements
corresponding to edges in G. If we restrict $\bfK$ to the space $\cone
^+(G) \cap\mathcal{S}^{0}$, we have the following constraints on the
off-diagonal elements of the Cholesky square root $\bfPhi$:
\begin{align}
\Phi_{ii} & > 0 \text{ for } i = 1, \dots, n, \label{diagfree} \\
\Phi_{ij} & = -\frac{1}{\Phi_{ii}} \sum_{d=1}^{i-1} \Phi_{di}\Phi_{dj}
\text{ for } (i,j) \not\in E, \label{completion}\\
\Phi_{ij} & < -\frac{1}{\Phi_{ii}} \sum_{d=1}^{i-1} \Phi_{di}\Phi_{dj}
\text{ for } (i,j) \in E. \label{myineq}
\end{align}
The first two conditions guarantee that $\bfPhi^T\bfPhi \in\cone
^+(G)$. The addition of the third inequality guarantees that $\bfPhi^T
\bfPhi\in\mathcal{S}^0$; however, this restriction comes at the cost
of losing variation independence (i.e., the parameters space of $\bfPhi
$ is no longer rectangular).

\subsection{Sampling from the Truncated G-Wishart Distribution}
We sample from the truncated G-Wishart distribution using a random
walk Metropolis--Hastings algorithm similar to the sampler proposed by
\cite{dobra2011bayesian}. We sequentially perturb one free element $\Phi
_{i_0j_0}$ at a time, holding the other free elements constant. In
doing so, we must find the support of the conditional distribution of
$\Phi_{i_0 j_0}$ given the other elements. The support of this
conditional distribution is the set of $\Phi_{i_0 j_0}$ that satisfy
inequalities (\ref{diagfree})--(\ref{myineq}) when the free elements,
the left-hand sides of (\ref{diagfree}) and (\ref{myineq}), are fixed.

For each specific graph and fixed pair $(i_0, j_0)$, we can write the
inequalities in (\ref{myineq}) as
\begin{align}
\Phi_{ij} & < g_{ij}\left(\Phi_{i_0 j_0}, \mathcal{F}_{-(i,j)}\right)
\text{ for } (i,j) \in E, \nonumber
\end{align}
where $\mathcal{F}_{-(i,j)}$ is the set of fixed, free elements of
$\bfPhi$ excluding $\Phi_{ij}$ and $\Phi_{i_0 j_0}$. We construct
$g_{ij}$ by substituting the equalities from (\ref{completion}) for all
of the non-free elements that depend on $\Phi_{i_0 j_0}$. Each $g$ is
(at worst) a quadratic function of $\Phi_{i_0 j_0}$. When $g$ is a
linear function, solving $g$ for $\Phi_{i_0 j_0}$ gives a solution set
of the form $ g^{-1}_{ij} (\Phi_{ij}, \mathcal{F}_{-(i,j)} )
= \{\Phi_{i_0, j_0} \in(L_{ij}, \infty) \}$, where $L_{ij}
< 0$. When $g$ is quadratic, the solution set is $ g^{-1}_{ij}(\Phi
_{ij}, \mathcal{F}_{-(i,j)})= \{\Phi_{i_0 j_0} \in(L_{ij},
U_{ij})\}$, where $L_{ij}$ is again negative.

If $(i,j) \prec(i_0, j_0)$ in lexicographical order, then the upper
bound for $\Phi_{ij}$ cannot depend on $\Phi_{i_0 j_0}$. Depending on
the graphical structure, there are pairs $(i,j) \succ(i_0, j_0)$ such
that the bound for $\Phi_{ij}$ does not depend on $\Phi_{i_0 j_0}$. In
these cases $g^{-1}_{ij}(\Phi_{ij}, \mathcal{F}_{-(i,j)} ) =
(-\infty, \infty)$.

\begin{theorem}
The conditional distribution of a free element $\Phi_{i_0 j_0}, \, i_0
\neq j_0$ given all other free elements is a continuous distribution
over an open subinterval of $\mathbb{R}^-$ given by
\begin{align}
\bigcap_{(i,j) \in E} g^{-1}_{ij}\left(\Phi_{ij}, \mathcal{F}_{-(i,j)}
\right) \cap\left(-\infty, \frac{-1}{\Phi_{i_0i_0}} \sum_{d=1}^{i_0-1}
\Phi_{di_0}\Phi_{dj_0} \right). \nonumber
\end{align}
\end{theorem}

We now give the analogous theorem for free, diagonal elements:
\begin{theorem}
The conditional distribution of a free element $\Phi_{i_0 i_0}$ given
other free elements is a continuous distribution over a subinterval of
$\mathbb{R}^+$ given by
\begin{align} \nonumber
\Phi_{i_0 i_0} & \in\left( \max_{i_0< k \leq p, (i_0, k)\in E} \left\{
- \frac{\sum_{d=1}^{i_0-1} \Phi_{di_0} \Phi_{dk}}{\Phi_{i_0 k}} \right\}
, \infty\right) \text{ for $1<i_0<n$},\\
\Phi_{i_0 i_0} & \in(0, \infty) \text{ for $i_0=1,n$}.\nonumber
\end{align}
\end{theorem}
For proofs, see the supplementary material \citep{Smithetal2015}.

We use these bounds to construct a Markov chain with stationary
distribution equal to the truncated G-Wishart distribution. Suppose
$\bfPhi^t$ is an upper-triangular matrix at iteration $t$ such that
$(\bfPhi^t)^T\bfPhi^t \in \cone^+(G) \cap\mathcal{S}^0$. For each
free element in $\Phi_{i_0j_0}^t$ do the following:
\begin{enumerate}
\item Calculate the upper and lower limits for $ \Phi_{i_0j_0}^t$ as
described above.
\item Sample from a truncated normal with these limits, mean $\Phi
_{i_0j_0}^t$, and standard deviation $\sigma_{m}$.
\item Update the non-free elements in lexicographical order. These
steps give a proposal $\bfK'=(\bfPhi')^T \bfPhi'$ where the free
elements in $\bfPhi'$ equal to the free elements of $\bfPhi^t$ except
in the $(i_0, j_0)$ entry.
\item Accept according to the acceptance probability $\alpha= \min(1,
R_m)$, where
\begin{align}
R_m & = \frac{\pi(\bfK'\mid\bfD, \delta, G)q(\bfK^t\mid\bfK')}{\pi
(\bfK^t\mid\bfD, \delta, G)q(\bfK'\mid\bfK^t)} \nonumber\\
& = \left(\frac{\Phi'_{i_0i_0}}{\Phi^t_{i_0i_0}} \right)^{\delta+\nu
_i(G)-1} \exp\left(-\frac{1}{2} \left<\bfK'-\bfK^t,\bfD\right>\right)\nonumber\\
&\quad{}
\times\frac{\text{TNorm}(\Phi^t_{i_0j_0};\Phi'_{i_0j_0}, \sigma_m, l_{i_0
j_0},u_{i_0j_0})}{\text{TNorm}(\Phi'_{i_0j_0};\Phi^t_{i_0j_0}, \sigma
_m, l_{i_0j_0}, u_{i_0j_0})}. \nonumber
\end{align}
\end{enumerate}
$\text{TNorm}(\cdot; \mu, \sigma, l, u)$ is the density of a normal
distribution with mean $\mu$ and standard~deviation $\sigma$ truncated
to the interval $(l, u)$, and $\nu_i(G)$ is the number of areas that
are neighbors of area $i$ but have larger index numbers, that is, $\nu
_i(G) = \# \{ j : \omega_{ij} = 1 \text{ and } i<j \}$.

\begin{figure}[h]
\includegraphics{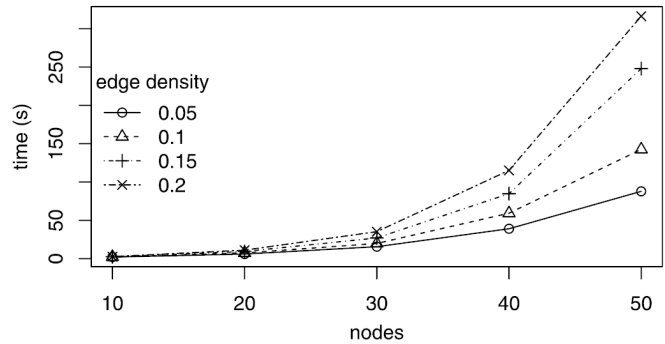}
\caption{Time to 1000 iterations by edge density and number of nodes.
For reference, the edge density of the counties in Washington State is
$0.123$, and the edge density of the continental US and the District of
Columbia is $0.093$.}
\label{fig:scale}
\end{figure}

The speed of this sampler depends on both the number of areas and on
the number of edges in the adjacency graph. These determine the number
of non-zero elements in $\bfK$ and the number of non-zero elements in
$\boldsymbol{\Phi}$. The elements of $\bfK$ can be reordered to form a
banded matrix. The size of the bandwidth depends on the proportion of
non-missing edges (i.e., the edge density), and the bandwidth of
$\boldsymbol{\Phi}$ is the same as $\bfK$ \citep{rue2005gaussian}. Thus
reordering the elements of $\bfK$ can create sparsity in $\boldsymbol
{\Phi}$, which reduces the number of nonzero terms in \eqref{myineq}.
Figure \ref{fig:scale} shows the time to one thousand iterations for
graphs with different numbers of nodes and edges, averaging over 50
simulated networks for each size-density combination. For each
simulation, we randomly sample networks with a given size and density
and reorder the elements using a bandwidth-decreasing algorithm (the
reverse Cuthill--McKee algorithm, available in the \texttt{spam}
package). The sampler scales well for very sparse networks, but the
time to 1000 iterations grows quickly when the edge density is over
$20$\%. The edge densities of the counties in Washington State and the
states in the continental US are $0.123$ and $0.093$,
respectively.\looseness=-1
%

\subsection{Using the Truncated G-Wishart in a Hierarchical Model}
We use truncated G-Wishart prior within the generic Bayesian
hierarchical model for areal counts given in Section 2:
\begin{align*}
\log(\theta_i) & = \bfx_i^T\beta+ u_i,\\
\pi(\mathbf{u} \mid\alpha, \tau_u, \bfK) & = \normal\left( \alpha
\mathbf{1}, (\tau^2_u \bfK)^{-1} \right),\\
\pi(\alpha) & = \normal(0, \sigma^2_\alpha),\\
\pi(\beta) & = \normal(0, \sigma^2_\beta\mathbf{I}),\\
\pi(\tau_u^2\mid a, b) & = \mathsf{Gam}(a, b),\\
\pi\left(\bfK\mid G,\delta, \bfD\right)& = \mathsf{TWis}_G\left(\delta
, (\delta- 2) \bfD(\rho)\right) \text{ with } K_{11} = W_{1+},\\
& \,\,\,\,\,\,\,\bfD^{-1}(\rho) = \bfD_W-\rho W,\\
\pi(\rho) &= \text{Unif}(0,0.05,0.1,\dots,\\
& \,\,\,\,\,\,\, 0.8,0.82,\dots,0.90,0.91,\dots,0.99).
\end{align*}
We suggest choosing the hyper parameters for the priors on $\alpha$ and
$\tau^2$ by first specifying a reasonable range for the average
relative risk and then finding values of $\sigma^2_\alpha$ and $(a,b)$
that match this range for a fixed value of $\bfK$. For fixed $\bfK$,
the distribution of $\overline{\bfu} = 1/n \sum_{i=1}^n u_i$ is a
univariate normal distribution depending on $\alpha$ and $\tau^2$.
Using the adjacency matrix of Washington State as an example and
letting $\bfK= \bfD^{-1}(0.99)$, $95\%$ of the prior on $\exp(\overline
{\bfu})$ is between $(1/8, 8)$ when $\sigma^2_\alpha= 1$ and $(a,b) =
(0.5, 0.0015)$. For a more informative prior, setting $\sigma^2_\alpha
= 1/4$ gives a range of $(1/2, 2)$. More details of this prior
specification framework are in the supplementary material.

The prior on the spatial autocorrelation parameter $\rho$ was
introduced by \cite{gelfand2003proper} for computational convenience
and to reflect the fact that large values of $\rho$ are needed to
achieve non-negligible spatial dependence in the proper CAR prior. \cite
{jin2007order} use a continuous uniform prior on $(0,1)$ and a $\mathsf
{Beta}(18,2)$ prior in a similar multivariate context. For our
purposes, using a discrete prior for $\rho$ is essential for carrying
out MCMC because $\rho$ appears in the normalizing constant of the
prior on $\bfK$. That is, the normalizing constant in (\ref
{ngwishdistro}) becomes $I_2(G, \delta, \bfD(\rho))$. As will be shown
below, we pre calculate ratios of these normalizing constants in
advance. It is not practical to repeat this process at each step of the MCMC.

We estimate the posterior distribution of the relative risks,
$\boldsymbol{\theta}$, using MCMC. Most of the transitions are standard
Metropolis or Gibbs updates (see supplementary material) except for the
updates on the precision matrix $\bfK$ and the autocorrelation
parameter $\rho$. We update $\bfK$ as described in Section 3.3,
skipping over $\Phi_{11}$ to preserve the restriction on $K_{11}$. We
update $\rho$ by choosing the next smallest or largest value in $\{
0,0.05,0.1,\dots,0.8,0.82,\dots,$ $0.90,0.91,\dots,0.99\}$, each with
probability $1/2$. If $\rho_t$ and $\rho'$ are not on the boundary of
this list, then the acceptance probability is $\alpha_\rho= \min(1,
R_m)$ where
\begin{align}
\log(R_m) & = -1/2 \text{tr}\left[(\delta-2)\bfK\left\{ (\bfD_w - \rho
' \bfW)^{-1} - (\bfD_w - \rho_t \bfW)^{-1} \right\}\right] \label
{acceptrho} \\
& \,\,\,\,\,\,\,\,\,\,\,+ \log\left[I_2\left(G, \delta, (\delta- 2)
\bfD(\rho_t)\right)\right] - \log\left[I_2\left(\delta, (\delta- 2)
\bfD(\rho')\right)\right]. \nonumber
\end{align}
If either $\rho_t$ or $\rho'$ is on the boundary, there is an extra
factor of $2$ because the proposal is not symmetric: if $\rho_t = 0$,
we propose $\rho' = 0.05$ with probability $1$. Because the graph $G$
is constant, the normalizing constants in (\ref{acceptrho}) only depend
on $\rho$. We estimate the necessary ratios of normalizing constants
and store them in a table prior to running the full MCMC.

For two densities of the form $\pi_1(\eta) = c_1 q_1(\eta)$ and $\pi
_2(\eta) = c_2 q_2(\eta)$ with normalizing constants $c_1$ and $c_2$,
the ratio of normalizing constants is given by $r = c_1/c_2 = \E_2
[q_1(\eta)/q_2(\eta)]$ when the support of the two distributions
are the same \citep{ming2000monte}. Here $\E_2$ is the expectation
under the second density. We estimate this expectation for each
consecutive pair $\rho_1 > \rho_2$ using MCMC. Here we give the details
for estimating the normalizing constants of a set of G-Wishart
distributions without restrictions on the $K_{11}$ element and with
$\delta= 3$. However, the same process will work for the truncated
G-Wishart and with the restriction that $K_{11} = W_{1+}$.
\begin{itemize}
\item Generate a Markov chain $\bfK_1, \, \bfK_2, \, \dots,\bfK_S$ with
stationary distribution\break  $\mathsf{Wis}_G(3, (\bfD_w - \rho_2 \bfW)^{-1})$.
\item For each state, let $Z_i = -1/2 \text{tr}[\bfK_i ( (\bfD
_w - \rho_1 \bfW)^{-1} - (\bfD_w - \rho_2 \bfW)^{-1} ) ]$.
\item Estimate $\log[I_1(G, 3, \bfD(\rho_1))] - \log[I_1(G,3, \bfD(\rho_2))]$ by $\log[ \frac{1}{S} \sum_{i =
1}^S \exp( Z_i) ]. $
\end{itemize}
%
For each pair $(\rho_1, \rho_2)$, we average over the estimates from
$10$ parallel chains of $100{,}000$ iterations. Figure 6 in the
supplementary material shows the evolution of the estimates of
$\log[ I_1(G, 3, (\bfD_w - 0.99 \bfW)^{-1})]-
\log[I_1(G,3, (\bfD_w - 0.98 \bfW)^{-1} )]$ using
the adjacency graph of the counties in Washington State.

\subsection{Multivariate Disease Mapping}
In Section 5, we use the truncated G-Wishart prior to analyze incidence
data from the Washington State Cancer Registry. In doing so, we adopt
the same framework as \cite{dobra2011bayesian} and assign a matrix
normal prior with a separable covariance structure to the log relative
risks. This means we assume that the covariance in the log relative
risks factors into a purely spatial portion and a purely
between-outcomes portion.
This assumption is common for modeling two-way data including
multivariate spatial data \citep{gelfand2003proper,carlin2003hierarchical,jin2007order} and spatio-temporal data
\citep{knorr2000insep,stein2005space,quick2013modeling} as well as
multi-way data \citep{mardia1993spatial,fosdick2014separable}.

Here we assume that there are $n$ areas with counts for $C$ cancer
sites (site of primary origin of the cancer) observed in each area. If
$\mathbf{Y} = \{y_{ic}: i = 1, \dots, n, \, c = 1, \dots, C \}$ is a
matrix of observed counts and $\mathbf{E} = \{E_{ic}: i = 1, \dots, n,
\, c = 1, \dots, C \}$ is a matrix of expected counts, then we have
\begin{align*}
y_{ic}| E_{ic}, \, \theta_{ic} & \sim\Poi\left(E_{ic} \theta_{ic}\right
), \\
\log(\bfTheta) & = \bfU,\\
\bfU& \sim\MN\left(\mathbf{M}, \bfK_{C}^{-1}, \bfK_{R}^{-1} \right),\\
M_c & \sim\normal\left(0, \sigma^2_M\right) \text{ for }c = 1, \dots
,C,\\
\bfK_{C} & \sim\mathsf{Wis}\left(\delta_C, (\delta_C-2) \mathbf{I}
\right) \text{ or } \mathsf{Wis}_{G_C}\left(\delta_C, (\delta_C-2)
\mathbf{I}\right),\\
\bfK_{R} & \sim\mathsf{TWis}_{G_R}\left( \delta_R, (\delta_R-2) \bfD
_R^{-1} \right).
\end{align*}
We use $\MN(\mathbf{M}, \bfSigma_C, \bfSigma_R)$ to denote the matrix
normal distribution with separable covariance structure \citep
{dawid1981some}. That is $\mbox{vec}( {\bfU})\mid\mathbf
{M},\bfSigma_{R},\bfSigma_{C} \sim\normal( \mbox{vec}\{
\mathbf{M} \},\bfSigma_{C} \otimes\bfSigma_{R})$, where
``$\otimes$'' is the Kronecker product. In the absence of any
information on cancer risk factors such as smoking rate or a
socioeconomic summary measure, we only include an overall rate for each
cancer in the mean model, that is, $M_{ic} = M_c$. The row covariance
$\bfSigma_R$ describes the spatial covariance structure of the log
relative risks. The column covariance matrix $\bfSigma_C$ describes the
covariance between the cancers.

We incorporate the truncated G-Wishart distribution as the prior for
the spatial precision matrix $\bfSigma_R^{-1} = \bfK_R$, and we use a
G-Wishart or Wishart prior with mode equal to the identity matrix for
$\bfSigma_C^{-1} = \bfK_C$. When the prior on $\bfK_C$ is a G-Wishart
prior, we incorporate uncertainty in the between-cancer conditional
independence graph $G_C$ using a uniform prior over all graphs. For
both priors, we restrict $(\bfK_C)_{11} = 1$ for identifiability.
Finally, we use an independent normal prior on each $M_c$. We estimate
the relative risks under this model using an MCMC sampler identical to
that in \cite{dobra2011bayesian}, substituting in the sampler from
Section 3.2 for the update on $\bfK_R$.

The assumption of separability yields a more parsimonious covariance
structure and can yield more stable estimation than with a full,
unstructured covariance matrix. Conditioning on one precision matrix
forms `replicates' for estimating the other:
\begin{align*}
\text{vec}(\mathbf{U}) \sim \mathsf{N}(0, \mathbf{K}_C^{-1} \otimes
\mathbf{K}_R^{-1} ) \implies (\mathbf{I}_{C} \otimes\boldsymbol{\Phi
}_R) \cdot\text{vec}(\mathbf{U}) & \sim\mathsf{N}(0, \mathbf
{K}_C^{-1} \otimes\mathbf{I}_R ).
\end{align*}
Thus, if $\bfK_C$ is known, then the sample size for estimating $\bfK
_R$ is equal to the number of rows, and similarly, if $\bfK_R$ is
known, the sample size for estimating $\bfK_C$ (and $G_C$) is equal to
the number of columns. This factorization appears in the iterative
algorithm for finding the maximum likelihood estimates of the matrix
normal distribution \citep{dutilleul1999mle} as well as in the Gibbs
sampler when using the conjugate Wishart prior with matrix or array
normal data \citep{hoff2011separable}.

\section[Simulation]{Simulation Study}
We compare the univariate disease mapping model using the truncated
G-Wishart prior to three other models in a simulation study based on a
similar study in \citet{lee2013bayesian}. The purpose of this
simulation study is to investigate the potential of the truncated
G-wishart prior in a Bayesian hierarchical model for a single
realization of a disease outcome in each area. We also directly compare
the G-Wishart to the standard Gaussian Markov random field formulation
in a univariate context, which has not previously been done in the
literature. We find that the more flexible G-Wishart priors can be
advantageous when the underlying disease risk surface has sharp
changes, but there are serious concerns related to estimating a large
number of covariance parameters \mbox{(39 + 93 = 132 for our example)}. We
illustrate a more realistic example relying on the assumption of
separability in Section 5.

\subsection{Data Generation}
\begin{figure}[t!]
\includegraphics{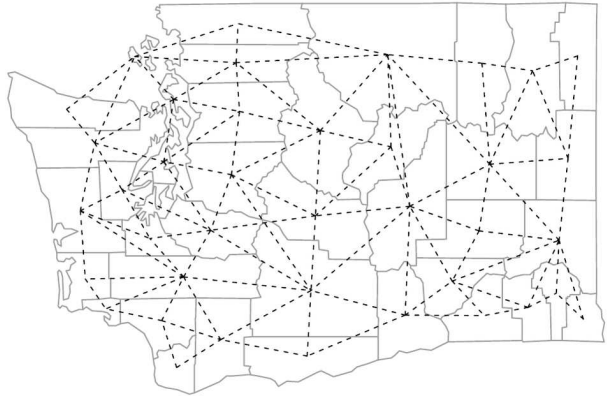}
\caption{Washington counties and adjacency graph:
39 areas, 93 edges, 648 missing edges.}
\label{adjacency}
\end{figure}

\begin{figure}[t!]
\includegraphics{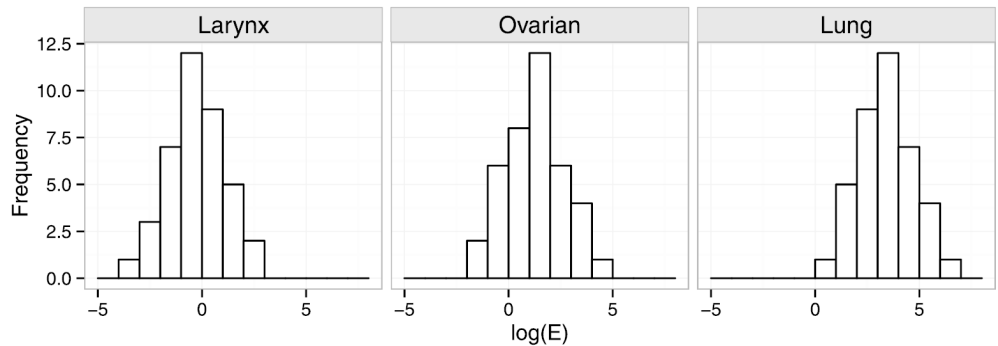}
\caption{Distribution of the log expected counts. Expected counts are
based on the 2010 population in each county and published rates for
larynx, ovarian, and lung cancer in the UK. These three cancers
represent a range of disease incidence from rare to common.}
\label{ExpCounts}
\end{figure}

We use the $39$ counties in Washington State as our study region and
generate expected counts based on the age-gender structure of these
counties in the $2010$ Census and published rates for larynx, ovarian,
and lung cancer in the United Kingdom in 2008 \citep{cancerstats}.
These three cancers are chosen to represent a range of disease
incidence from rare to common. A map of the counties with the
underlying undirected graph is shown in Figure \ref{adjacency}, and the
distributions of expected counts for each cancer are shown on the log
scale in Figure \ref{ExpCounts}.
%

%

\begin{figure}[t!]
\includegraphics{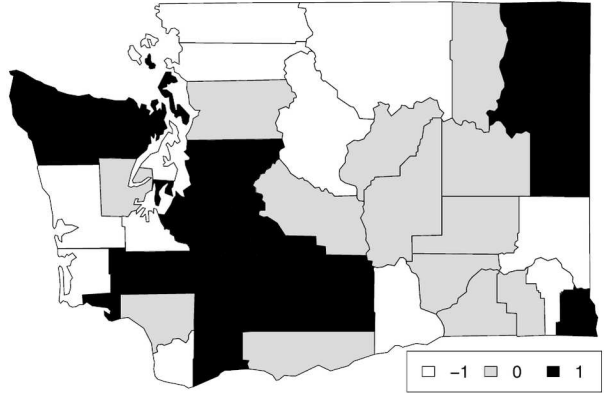}
\caption{Labels ($L_i$) for simulation study.}
\label{simlabs}
\end{figure}

We generate the risk surface as the combination of a globally-smooth
surface and a locally-constant surface. We label each area $-1,\,0\,$
or $1$ using a Potts model \citep{green2002hidden} so that neighboring
areas are more likely to have the same label. The label allocation for
this simulation study is shown in Figure \ref{simlabs}. For each
simulation, we generate
\begin{align}\nonumber
y_i & = \Poi(E_i \theta_i),\\
\log(\theta_i) & = 0.1 x_i + (M\times L_i + u_i), \nonumber
\end{align}
where $L_i$ is the label assigned to county $i$. We simulate $x_i$ and
$u_i$ independently from multivariate normal distributions with Mat\'
{e}rn covariance function with smoothness parameter $2.5$ and range
chosen so that the median marginal correlation is $0.5$. Thus, each of
the vectors $\bfx$ and $\bfu$ are realizations of a smooth spatial
process observed at a finite set of points. In different simulations,
we set $M$ to $0.5$, $1$, or $1.5$. Larger values of $M$ lead to a risk
surface with more discontinuities. We generate $50$ realizations from
each combination of $M$ and the three sets of expected counts.
%

For the simulation results described below, we run each chain for
$100{,}000$ iterations, discarding the first half as burn in. We set the
prior parameters for the model in Section~3.3 to $\sigma_\alpha= 1$,
$\sigma_\beta= 10$, $(a,b) = (0.5,0.0015)$, and $\delta= 3$. Figure 3
in the supplementary material shows the evolution of the posterior mean
for $10$ different chains for two elements of the Cholesky square root
and two random effects. In all cases, we reach convergence in about
$10{,}000$ iterations.

\subsection{Results}
We compare the model using the truncated G-Wishart prior to three other
models. The model using the G-Wishart prior is identical to the model
from Section 3.3 except that the prior on the precision matrix $\bfK$
is the G-Wishart prior instead of the truncated G-Wishart prior. We
also compare against the convolution model from Section 2.2 and a
similar model that includes only spatial random effects with an ICAR
prior. In the convolution and ICAR models, we estimate the posterior
mean and variance of the relative risks using INLA. For the models
using truncated G-Wishart and G-Wishart priors, we explore the
posterior distributions using MCMC.

\begin{figure}
\includegraphics{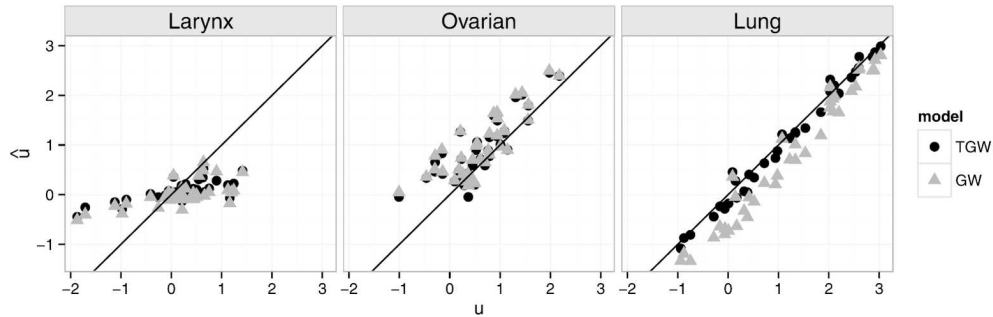}
\caption{Simulated versus estimated spatial random effects for one
simulation from each of set of expected counts with M = 0.5. The
estimates are the posterior means of $\mathbf{u}$ under the truncated
G-Wishart (TGW) and G-Wishart (GW) models. The posterior estimates
shrink toward the prior mean of zero
as the expected counts decrease.}\label{fig:shrinkTGWGW}
\end{figure}

In Figure \ref{fig:shrinkTGWGW}, we compare the true spatial random
effects $\bfu$ against the posterior estimates of the random effects
for the truncated G-Wishart model and the G-Wishart models from one
simulation for each set of expected counts. The estimates of the random
effects are similar to the true values when the expected counts are
high, but there is substantial shrinkage toward the prior mean of zero
when the expected counts are small. This reflects the fact that there
is much more information about the relative risks when the counts $\bfy
$ are larger, and we see the same relationship in other disease mapping models.

%

\begin{figure}[p]
\includegraphics{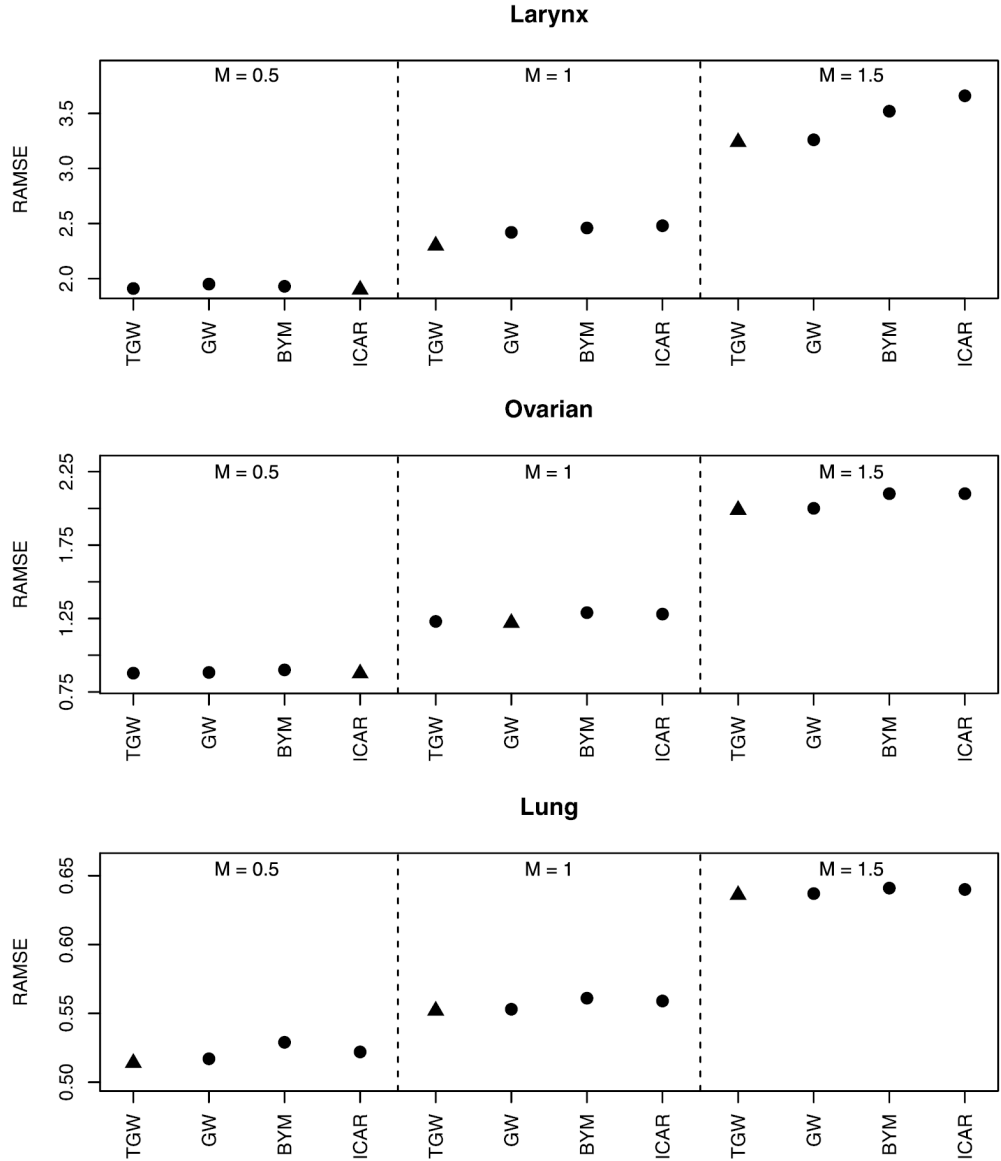}
\caption{Root average mean squared error (RAMSE) for relative risks
$\boldsymbol{\theta}$. The triangle signifies the smallest value for
each experiment. The four models are: TGW, truncated G-Wishart prior on
the precision matrix for the spatial random effects; GW, G-Wishart
prior on the precision matrix for the spatial random effects; BYM,
convolution model with independent and ICAR random effects; ICAR, only
ICAR random effects. All models show increased RAMSE with increased
spatial discontinuities (large M) and increased RAMSE with smaller
expected counts. The TGW prior performs the best in six out of nine
scenarios with the greatest benefit in the larynx, $M=1.5$ experiment.}
\label{simres}
\end{figure}

We compare the four methods using the root-averaged mean squared error
(RAMSE) of the posterior mean of each relative risk $\theta_i$. This is
the square root of the mean squared error averaged over all simulations
and all areas. For $S$ simulations and $B$ iterations of the MCMC
sampler, the RAMSE is
\begin{align*}
\text{RAMSE} = \sqrt{\frac{1}{39\times S \times B}\sum_{i =1}^{39} \sum
_{s=1}^{S} \sum_{b=1}^B (\theta_{is}^{(b)} - \theta_{is})^2 },
\end{align*}
where $\theta_{is}$ it the true relative risk for area $i$ in
simulation $s$ and $\theta_{is}^{(b)}$ is the corresponding value at
iteration $b$ of the MCMC. The results of this simulation are shown in
Figure \ref{simres}, and the triangle indicates the lowest RAMSE within
each scenario.

In general, the RAMSE decreases for all four models when the expected
counts increase, and the RAMSE increases when the level of smoothing
decreases (i.e.,~M~increases). The model using the truncated G-Wishart
prior performs the best in six out of nine scenarios, and we see the
greatest benefit in the larynx, $M=1.5$ simulation when the expected
counts are low and the local discontinuities in the risk surface are
most prominent.

%

%
\begin{figure}[t!]
\includegraphics{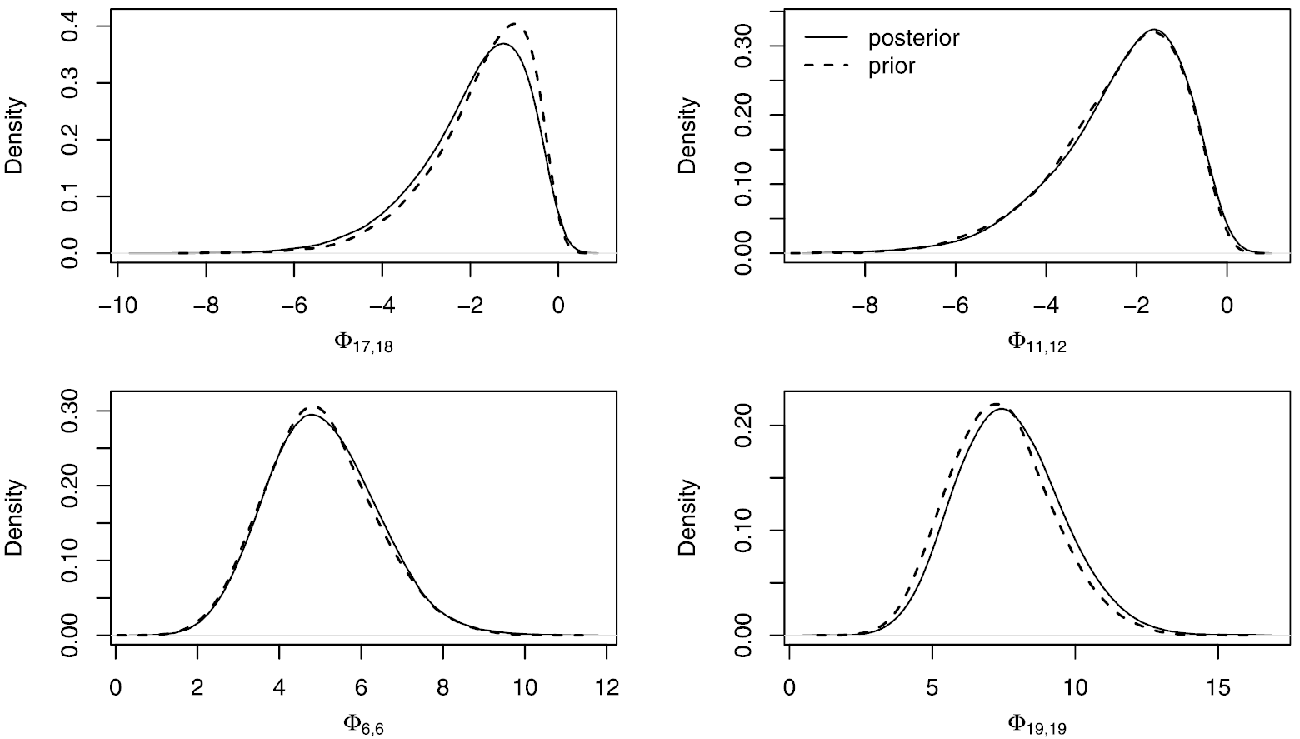}
\caption{Comparison of the prior versus the posterior distribution of
the elements of the Cholesky square root for one realization of the
univariate simulation study with the TGW model.}
\label{fig:prvpostUniv}
\end{figure}

While the truncated G-Wishart and G-Wishart priors for the spatial
covariance appear advantageous in this simulation study, there is
little information in a single sample for estimating the full
covariance matrix. Figure \ref{fig:prvpostUniv} shows that the
posterior distributions of the elements of the Chokesly square root are
nearly identical to the prior distributions. This suggests that prior
parameter choice plays a substantial role in the results from the TGW
and GW models. Furthermore, the TGW and GW models should struggle when
the risk surface is smoothly varying and the degree of smoothness is
common across the study region. Table 2 in the supplementary material
shows that the convolution model outperforms the TGW and GW models when
there is no spatial association (the log relative risks are generated
independently) and when the underlying risk surface is smooth (the log
relative risks are generated directly from the ICAR prior). In general,
the TGW and GW results are comparable with the convolution\vadjust{\goodbreak} model when
the expected counts were larger and there was some spatial structure in
the risk surface. However, with large expected counts (e.g., the lung
cancer scenario), most reasonable methods will perform adequately.

\section[Multiple Disease Mapping]{Multiway Disease Mapping}
In this section, we use the truncated G-Wishart prior in a multivariate
disease mapping context using cancer incidence data from the Washington
State Cancer Registry. Let $\mathbf{Y} = \{y_{ic}: i = 1, \dots, 39, \,
c = 1, \dots, 10\}$ be a $39\times10$ matrix of incidence for $10$
cancers in each county in Washington State in 2010. These $10$ cancers
have the largest incidence across the state in 2010. The expected
counts $E_{ic}$ are calculated separately for each cancer using
internal standardization based on sex and 5-year age bands. The
standardized incidence ratios (SIRs = $\bfY/ \bfE$) for these data are
between $0$ and $3.91$, and the range of the empirical correlations
between the SIRs of the different cancers (not~taking into account
spatial dependence) is $(-0.203, 0.477)$. Just over $20\%$ of the
counts are under~$5$, but we do not treat small counts as missing in
this analysis.

We use cross-validation to compare the model in Section 3.4 to models
using the G-Wishart prior \citep{dobra2011bayesian} and using the
proper CAR form for $\bfK_R$ \citep{gelfand2003proper}. We compare $3$
different choices for the prior on $\bfK_R$ and two choices for the
prior on $\bfK_C$. For the truncated G-Wishart and G-Wishart priors on
$\bfK_R$, we set $\delta_R = 3$ and $\bfD_R=\bfD(\rho) = (\bfD_\omega-
\rho\bfW)^{-1}$, where the prior on $\rho$ is the same as in Section~3.3.
The MCAR prior on $\bfK_R$ is simply $\bfK_R = \bfD(\rho)^{-1}$.
For both the Wishart and the G-Wishart priors on $\bfK_C$, we set
$\delta_C = 3$ and $\bfD_C = \bfI$.

We randomly split all observations into $10$ bins and create $10$ data
sets, each with one bin of counts held out. We impute the missing
counts as part of the MCMC and compare the models based on average
predictive squared bias ($\text{BIAS}^2$) and average predictive
variance (VAR). Let $E_\mathcal{M}( Y_{ic})$ be the predicted value
under model $\mathcal{M}$, $\text{var}_\mathcal{M} (Y_{ic})$ be the
variance of the posterior predictive distribution, and $Y_{ic}$ be the
observed count. The comparison criteria are
\begin{align*}
\text{BIAS}^2_{\mathcal{M}} & = \frac{1}{39\times10} \sum_{Y_{ic}}
\left(E_\mathcal{M} (Y_{ic}) - Y_{ic}\right)^2,\\
\text{VAR}_{\mathcal{M}} & = \frac{1}{39 \times10} \sum_{Y_{ic}} \text
{var}_\mathcal{M} (Y_{ic}).
\end{align*}

The results (based on running each MCMC for 200,000 iterations) are
given in Table \ref{comparison}. The truncated G-Wishart model with a
G-Wishart prior on $\bfK_C$ performs best in terms of bias, and the
truncated G-Wishart model with a Wishart prior on $\bfK_C$ performs
best in terms of predictive variance. Using the truncated G-Wishart
prior for the spatial precision matrix improves over the G-Wishart
prior for both choices of prior for $\bfK_C$. The MCAR model is the
second best model in terms of MSE (the sum of $\text{BIAS}^2$ and VAR).

%
\begin{table}[t!]
\centering
\begin{tabular}{rrrr|cc}
\hline
$\times10^5$ & $\text{BIAS}^2$ & VAR & MSE& $\pi(\bfK_C)$ & $\pi(\bfK
_R)$ \\
\hline
GGM& 2.18 & 1.06 & 3.23 &G-Wis&G-Wis\\
TGGM & \textbf{1.25} & 0.73 & \textbf{1.98} &G-Wis&NG-Wis\\
\hline
FULL & 2.40 & 0.99 & 3.39 &Wis&G-Wis\\
TFULL & 1.61 &\textbf{ 0.69} & 2.29 &Wis&NG-Wis\\
\hline
MCAR & 1.31 & 0.82 & 2.13 &Wis&CAR\\
\hline
\end{tabular}
\caption{Ten-fold cross-validation results for the Washington State
cancer incidence data. The five models use the matrix normal random
effects model from Section 3.4. The priors on the precision matrices
are: GGM, G-Wishart priors on $\bfK_R$ and $\bfK_C$; TGGM, truncated
G-Wishart prior on $\bfK_R$ and G-Wishart prior on $\bfK_C$; FULL,
G-Wishart prior on $\bfK_R$ and Wishart prior on $\bfK_C$; TFULL,
truncated G-Wishart prior on $\bfK_R$ and Wishart prior on $\bfK_C$;
MCAR, proper CAR prior on $\bfK_R$ and Wishart prior on $\bfK_C$. In
the GGM and TGGM models, the cancer conditional independence graph
$G_C$ is random. In the other three models, $G_C$ is a complete graph.}
\label{comparison}
\end{table}

%
\begin{figure}[t!]
\centering
\includegraphics{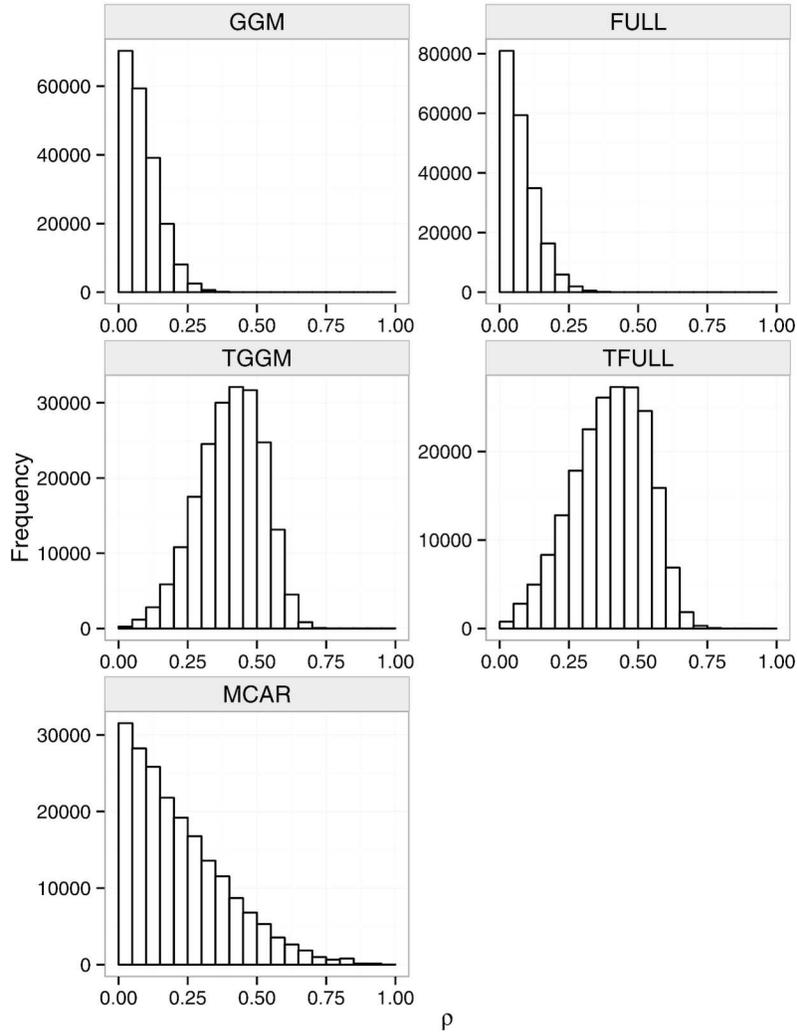}
\caption{Posterior distribution of the spatial autocorrelation
parameter $\rho$ under the five models considered.}
\label{autocor}
\end{figure}

%
\begin{figure}[t!]
\centering
\includegraphics{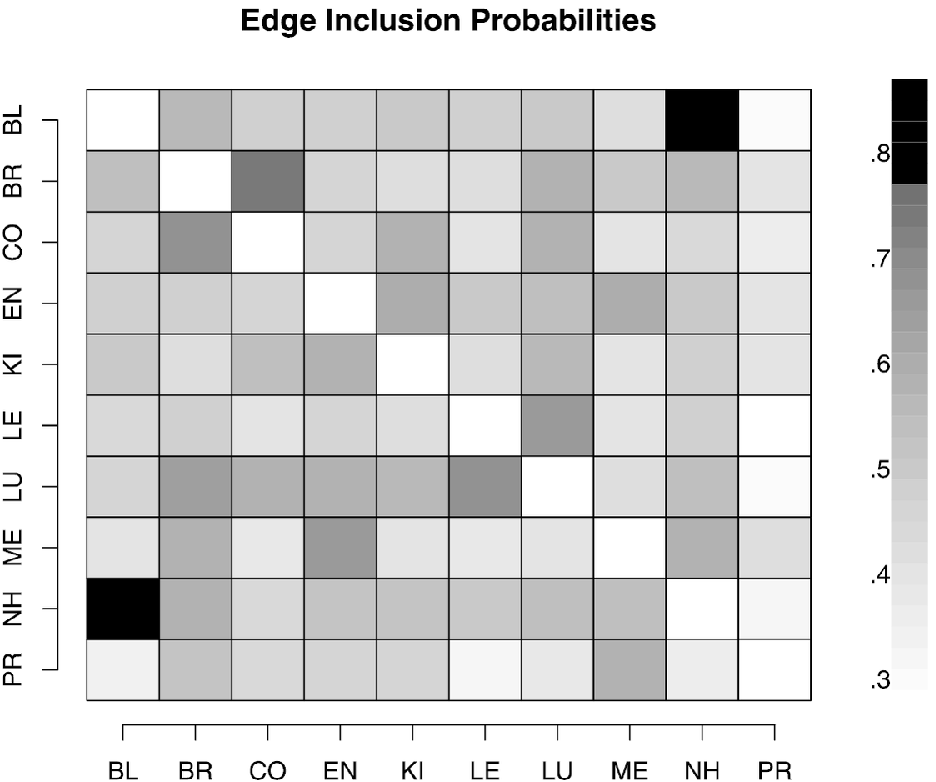}
\caption{Pairwise edge inclusion probabilities for $G_C$ when the prior
on $\bfK_R$ is G-Wishart (upper triangle) or truncated G-Wishart (lower
triangle). The abbreviations are: BL, Bladder; BR, Breast; CO,
Colorectal; EN, Endometrial; KI, Kidney; LE, Leukemia; LU, Lung; ME,
Melanoma of the skin; NH, Non-Hodgkin lymphoma; PR, Prostate. The
Lung--Leukemia, Bladder--Non-Hodgkin lymphoma, and Colon--Breast cancer
edges have the biggest posterior edge inclusion probabilities in both models.}
\label{graphs}
\end{figure}

Figure \ref{autocor} shows the estimated posterior distribution of the
spatial autocorrelation parameter $\rho$ for the five models. Under the
G-Wishart prior on $\bfK_R$, the posterior for $\rho$ is much more
concentrated near zero than with a truncated G-Wishart prior
(regardless of the prior on $\bfK_C$). The posterior median for $\rho$
when using CAR prior on $\bfK_R$ is between the estimates from the
G-Wishart and truncated G-Wishart priors. Figure \ref{graphs} shows the
estimated posterior probabilities of including edges in $G_C$ for two
different priors on~$\bfK_R$. The upper and lower triangles are quite
similar, indicating that inference on the between-cancer conditional
independence graph is not sensitive to the choice of prior on $\bfK_R$.
The Lung--Leukemia, Bladder--Non-Hodgkin lymphoma, and Colon--Breast
cancer edges have the biggest posterior edge inclusion probabilities.

Finally, we compare the GGM, TGGM, and MCAR models using within-sample
fit for the complete data. Table \ref{cov} shows the average coverage
and length of $95\%$ posterior predictive intervals as well as two
measures of the effective number of parameters: $p_{\text{DIC}}$ \citep
{spiegelhalter2002bayesian} and $p_{\text{WAIC}}$
\citep{gelman2013bayesian}:\vadjust{\goodbreak}
\begin{align*}
p_{\text{DIC}} & = 2\left( \log p (\bfY\mid\hat{\bf{\theta}}_{\text
{post}}) - \E_{\text{post}} \log p(\bfY\mid\bfTheta) \right),\\
p_{\text{WAIC}} & = \sum_{i = 1}^n \sum_{c = 1}^C \text{VAR} \log
p(Y_{ic} \mid\theta).
\end{align*}
While all models have approximately the correct coverage, the posterior
predictive intervals from the truncated G-Wishart model are slightly
smaller. This remains true when averaging over the predictive intervals
for small counts ($\leq5$) or larger counts ($\geq20$). The MCAR
model has the fewest number of effective parameters by both measures,
which is consistent with the parsimonious form of the spatial
covariance in the MCAR model. The G-Wishart model has the largest
number of effective parameters under $p_{\text{DIC}}$ but the truncated
G-Wishart has the largest number under $p_{\text{WAIC}}$. This
inconsistency in the ordering is likely a result of differences in the
shapes of the posterior predictive distributions under the GGM and TGGM
models. Both $p_{\text{DIC}}$ and $p_{\text{WAIC}}$ measure the spread
in the log posterior predictive density, but the estimators are
affected differently by features such as longer tails.

%
\begin{table}[t!]
\centering
\begin{tabular}{rrrrrrr}
\hline
& COV & LEN & $\text{LEN}_{\leq5}$ & $\text{LEN}_{\geq20}$ &
$p_{\text{DIC}}$ & $p_{\text{WAIC}}$ \\
\hline
GGM & 0.959 & 31.33 & 7.47 & 51.82 &184.1 & 133.1\\
TGGM & 0.954 & \textbf{ 31.27} & \textbf{7.36} & \textbf{51.79} &
182.2 & 135.3 \\
MCAR & 0.956 & 31.31 & 7.41 & 51.85 & 181.3& 131.3\\
\hline
\end{tabular}
\caption{Coverage rates (COV) and mean length (LEN) of the in-sample
$95\%$ credible intervals. Mean lengths are also give by ranges of
observed counts. $p_{\text{DIC}}$ and $p_{\text{WAIC}}$ are two
measures of the effective number of parameters.}
\label{cov}
\end{table}

The cross-validation results are somewhat sensitive to the choice of
prior on $\rho$. We investigated fixing $\rho$ to $0.99$ or $0.9$ (the
mean of the $\mathsf{Beta}(18,2)$ prior used in \cite{jin2007order})
as well as using a discrete uniform prior on $\{0.05,0.1,\dots
,0.9,0.95,0.99\}$. In some cases, the predictive variance is
substantially smaller than the variance in Table~\ref{comparison}, but
this comes at the cost of greater bias. The best method in terms of
overall MSE is still the TGGM model where the prior on $\rho$ is
discrete uniform with additional values closer to $1$. Full
cross-validation results for the three additional priors on $\rho$ are
in the supplementary material.

\section{Discussion}
This article presents a novel extension of the G-Wishart prior for the
precision matrix of spatial random effects. In a simulation study, the
truncated G-Wishart prior is able to better estimate the relative risks
when the outcomes are rare (i.e., the expected counts are small) and
when the risk surface is not smooth. However, we found that there is
not enough information in a single outcome to estimate the spatial
correlation structure. The restriction of the G-Wishart prior was shown
to be advantageous when used in a multivariate disease mapping context
with incidence data from the Washington State Cancer Registry.

The multivariate model relies on the assumption of separability to
estimate the rich correlation structure by pooling information across
outcomes. The validity of the separability assumption has been
carefully considered for spatiotemporal applications \citep{stein2005space,fuentes2006testing}, and alternative, non-separable
space--time covariance models have been proposed for Gaussian processes
\citep{gneiting2002nonseparable,gneiting2010continuous} and Gaussian
Markov random fields \citep{knorr2000insep}. \cite{gelfand2003proper}
extend the MCAR to allow for different spatial autocorrelation
parameters for each outcome, yielding non-separable model that is still
relatively parsimonious, and \cite{jin2005generalized,jin2007order} further extend the MCAR paradigm by
including parameters that a directly represent the correlation between
different outcomes in neighboring areas. Ultimately, these MCAR
extensions still make an assumption similar to separability in that the
correlation between outcomes within a single areas is the same for all
areas.\looseness=1

As mentioned in Section 2.1, others have approached this problem by
directly altering the conditional independence structure \citep
{knorr2000bayesian,green2002hidden,lee2013locally,lee2013bayesian}.
Given that these models have been shown to outperform the traditional
convolution model in some scenarios and are fairly parsimonious, these
methods may be better for univariate outcomes than our TGW model. One
direction for future research is to incorporate the locally adaptive
CAR \citep{lee2013locally} in the matrix variate random effect
framework of Sections 3.4 and 5.

There are a number of computation issues when using the truncated
G-Wishart and G-Wishart priors. Each MCMC run for the univariate
truncated G-Wishart model in Section 4 takes approximately $1.5$ hours
to complete on a $2.5 $ GHz Intel Xeon E5-2640 processor, and, with the
exception of the MCAR model, the MCMC for each model in Section 5 takes
about $6.5$ hours to complete. In contrast, estimating the convolution
and ICAR models from Section 4 takes a matter of seconds in INLA. We
have found that the proposal variance for updates of the Cholesky
square (Section 3.3) and the random effects (see supplementary
material) must be chosen carefully to avoid poor convergence. In both
Sections $4$ and $5$, we used $s = 2$ for updating $\bfPhi$ and $s=
0.1$ for updating $\mathbf{u}$. While the computation time for the
models detailed here are not prohibitive, they may pose a challenge as
we extend to more complicated datasets, such as those including
multiple diseases in time and space.

R code for the simulation in Section 4 and C++ code for the analysis
in Section 5 are available at \url{http://www.lancaster.ac.uk/staff/smithtr/NGWSource.zip}.
Included here are the expected counts and labeling scheme for Section 4
and prototypical data for Section 5. A censored version of the data
used in Section~5 is available from
\url{https://fortress.wa.gov/doh/wscr/WSCR/Query.mvc/Query}.

\begin{supplement}
\stitle{Supplementary Material for ``Restricted Covariance Priors with Applications in Spatial Statistics''}
\slink[doi]{10.1214/14-BA927SUPP}
\sdatatype{.pdf}
\sfilename{Supplement-Jan-06-2015.pdf}
\end{supplement}

\bibliographystyle{ba}
%

\begin{acknowledgement}
TS and AD were supported in part by the National Science Foundation
(DMS 1120255). JW was supported by 2R01CA095994-05A1 from the National
Institutes of Health. The authors thank the Washington State Cancer
Registry for providing the cancer incidence data and the referees for
their helpful comments.
\end{acknowledgement}

\end{document}